\documentclass[runningheads]{llncs}
\usepackage{graphicx}
\usepackage{subfigure}
\usepackage{amsmath}
\usepackage{amssymb}
\usepackage{amsbsy}
\usepackage{bm}
\usepackage{float}
\usepackage{cite}
\usepackage{algorithmic}
\usepackage{courier}
\usepackage{xcolor}
\usepackage{listings}
\usepackage[hyphens]{url}
\expandafter\def\expandafter\UrlBreaks\expandafter{\UrlBreaks
      \do\a\do\b\do\c\do\d\do\e\do\f\do\g\do\h\do\i\do\j%
      \do\k\do\l\do\m\do\n\do\o\do\p\do\q\do\r\do\s\do\t%
      \do\u\do\v\do\w\do\x\do\y\do\z\do\A\do\B\do\C\do\D%
      \do\E\do\F\do\G\do\H\do\I\do\J\do\K\do\L\do\M\do\N%
      \do\O\do\P\do\Q\do\R\do\S\do\T\do\U\do\V\do\W\do\X%
      \do\Y\do\Z}
\usepackage{mathptmx}
\usepackage{helvet}
\usepackage{type1cm}
\usepackage{makeidx}
\usepackage{multicol}
\usepackage[bottom]{footmisc}
\newcommand{\md}{\,\mathrm{d}}
\newcommand{\defas}{\triangleq}

\newcommand{\nullS}[1]{\mathop{\mathrm{null}}\left\{#1\right\}}

\newcommand{\V}[1]{\ensuremath{\boldsymbol{#1}}}
\newcommand{\M}[1]{\ensuremath{\boldsymbol{#1}}}
\newcommand{\VT}[1]{\ensuremath{\boldsymbol{#1}}^{\textrm{T}}}
\newcommand{\MT}[1]{\ensuremath{\boldsymbol{#1}}^{\textrm{T}}}
\newcommand{\T}{\ensuremath{}^{\textrm{T}}}

\renewcommand{\P}[1]{\ensuremath{\mathsf{#1}}}

\newcommand{\dt}[1]{\ensuremath{\frac{\md #1}{\md t}}}

\usepackage{empheq}
\newcommand*\widefbox[1]{\fbox{\hspace{2em}#1\hspace{2em}}}

\begin{document}
\title{Detection of Derivative Discontinuities\\ in Observational Data}
%
%
\author{Dimitar Ninevski\orcidID{0000-0003-0101-8686} \and
	Paul O'Leary\orcidID{0000-0002-1367-8270}}
\authorrunning{D. Ninevski and P. O'Leary}
\institute{University of Leoben, A8700 Leoben, Austria
	\email{automation@unileoben.ac.at}\\
	\url{http://automation.unileoben.ac.at}}
\maketitle              
\begin{abstract}
	This paper presents a new approach to the detection of discontinuities in the n-th derivative of observational data. This is achieved by performing two polynomial approximations at each interstitial point. The polynomials are coupled by constraining their coefficients to ensure continuity of the model up to the (n-1)-th derivative; while yielding an estimate for the discontinuity of the n-th derivative. The coefficients of the polynomials correspond directly to the derivatives of the approximations at the interstitial points through the prudent selection of a common coordinate system. The approximation residual and extrapolation errors are investigated as measures for detecting discontinuity. This is necessary since discrete observations of continuous systems are discontinuous at every point. It is proven, using matrix algebra, that positive extrema in the combined approximation-extrapolation error correspond exactly to extrema in the difference of the Taylor coefficients. This provides a relative measure for the severity of the discontinuity in the observational data. The matrix algebraic derivations are provided for all aspects of the methods presented here; this includes a solution for the covariance propagation through the computation. The performance of the method is verified with a Monte Carlo simulation using synthetic piecewise polynomial data with known discontinuities. It is also demonstrated that the discontinuities are suitable as knots for B-spline modelling of data. For completeness, the results of applying the method to sensor data acquired during the monitoring of heavy machinery are presented.
\end{abstract}
\section{Introduction}
In the recent past \emph{physics informed data science} has become a focus of research activities, e.g.,~\cite{Owhadi2015Bayesian}. It appears under different names e.g., \emph{physics informed}~\cite{Raissi2019}; \emph{hybrid learning}~\cite{Saxena2019}; \emph{physics-based}~\cite{yaman2019selfsupervised}, etc.; but with the same basic idea of embedding physical principles into the data science algorithms. The goal is to ensure that the results obtained obey the laws of physics and/or are based on physically relevant features. Discontinuities in the observations of continuous systems violate some very basic physics and for this reason their detection is of fundamental importance. Consider Newton's second law of motion,
\begin{equation}
F(t) = \dt{} \left\{ m(t) \, \dt{}y(t) \right\} =
\dot{m}(t)\,\dot{y}(t) + m(t)\,\ddot{y}(t).
\end{equation}
Any discontinuities in the observations of $m(t)$, $\dot{m}(t)$, $y(t)$, $\dot{y}(t)$ or $\ddot{y}(t)$ indicate a violation of some basic principle: be it that the observation is incorrect or something unexpected is happening in the system. Consequently, detecting discontinuities is of fundamental importance in physics based data science. A function $s(x)$ is said to be $C^n$ discontinuous, if $s \in C^{n-1}\setminus C^n$, that is if $s(x)$ has continuous derivatives up to and including order $n-1$, but the $n$-th derivative is discontinuous. Due to the discrete and finite nature of the observational data, only jump discontinuities in the $n$-th derivative are considered; asymptotic discontinuities are not considered. Furthermore, in more classical data modelling, $C^n$ jump discontinuities form the basis for the locations of knots in B-Spline models of observational data~\cite{Wahba1990}.
\subsection{State of the Art}
There are numerous approaches in the literature dealing with estimating regression functions that are smooth, except at a finite number of points. Based on the methods, these approaches can be classified into four groups: local polynomial methods, spline-based methods, kernel-based methods and wavelet methods. The approaches vary also with respect to the available a priori knowledge about the number of points of discontinuity or the derivative in which these discontinuities appear. For a good literature review of these methods, see \cite{Gijbels2005}. The method used in this paper is relevant both in terms of local polynomials as well as spline-based methods; however, the new approach requires no a priori knowledge about the data.

In the local polynomial literature, namely in \cite{Horvath2002} and \cite{Spokoiny1998}, ideas similar to the ones presented here are investigated. In these papers, local polynomial approximations from the left and the right side of the point in question are used. The major difference is that neither of these methods use constraints to ensure that the local polynomial approximations enforce continuity of the lower derivatives. Additionally, it is not clear whether only co-locative points are considered as possible change points, or interstitial points are also considered. Furthermore, both papers use different residuals to determine the existence of a change point. The latter also focuses on optimal subset selection, which is not the focus of this paper.

In \cite{Qiu1998} on the other hand, one polynomial instead of two is used, and the focus is mainly on detecting $C^0$ and $C^1$ discontinuities. Additionally, the number of change-points must be known a-priori, so only their location is approximated; the required a-priori knowledge make the method unsuitable in real sensor based system observation.

In the spline-based literature there are heuristic methods (top-down and bottom-up) as well as optimization methods. For a more detailed state of the art on splines, see \cite{Dung2017}. Most heuristic methods use a discrete geometric measure to calculate whether a point is a knot, such as: discrete curvature, kink angle, etc, and then use some (mostly arbitrary) threshold to improve the initial knot set. In the method presented here, which falls under the category of bottom-up approaches, the selection criterion is based on calculus and statistics, which allows for incorporation of the fundamental physical laws governing the system, in the model, but also ensures mathematical relevance and rigour.
\subsection{The New Approach}
This paper presents a new approach to detecting $C^n$ discontinuities in observational data. It uses constrained coupled polynomial approximation to obtain two estimates for the $n^\text{th}$ Taylor coefficients and their uncertainties, at every interstitial point. These correspond approximating the local function by polynomials, once from the left $\P{f}(x,\V{\alpha})$ and once from the right $\P{g}(x,\V{\beta})$. The constraints couple the polynomials to ensure that $\alpha_i = \beta_i \,\,\, \text{for every}\, i \in [0 \ldots n-1]$. In this manner the approximations are $C^{n-1}$ continuous at the interstitial points, while delivering an estimate for the difference in the $n^\text{th}$ Taylor coefficients. All the derivations for the coupled constrained approximations and the numerical implementations are presented. Both the approximation and extrapolation residuals are derived. It is proven that the discontinuities must lie at local positive peaks in the extrapolation error. The new approach is verified with both known synthetic data and on real sensor data obtained from observing the operation of heavy machinery.
\section{Detecting $C^n$ Discontinuities}
Discrete observations $s(x_i)$ of a continuous system $s(x)$ are, by their very nature, discontinuous at every sample. Consequently, we will require some measure for discontinuity, with uncertainty, which provides the basis for a statistical hypothesis test.

The observations are considered to be the co-locative points, denoted by $x_i$ and collectively by the vector $\V{x}$; however, we wish to estimate the discontinuity at the interstitial points, denoted by $\zeta_i$ and collectively as $\V{\zeta}$. Using interstitial points, one ensures that each data point is used for only one polynomial approximation at a time. Furthermore, in the case of sensor data, one expects the discontinuities to happen between samples. Consequently the data is segmented at the interstitial points, i.e. between the samples. This requires the use of interpolating functions and in this work we have chosen to use polynomials.

Polynomials have been chosen because of their approximating, interpolating and extrapolating properties when modelling continuous systems: The Weierstrass approximation theorem \cite{Weierstrass1885} states that if $f(x)$ is a continuous real-valued function defined on the real interval $x \in [a, b]$, then for every $\epsilon > 0$, there exists a polynomial $p(x)$ such that for all $x \in [a, b]$, the supremum norm $\|f(x) - p(x)\|_{\infty} < \epsilon$. That is \emph{any} function $f(x)$ can be approximated by a polynomial to an arbitrary accuracy $\epsilon$ given a sufficiently high degree.

\begin{figure}[h]
	\centering
	\includegraphics{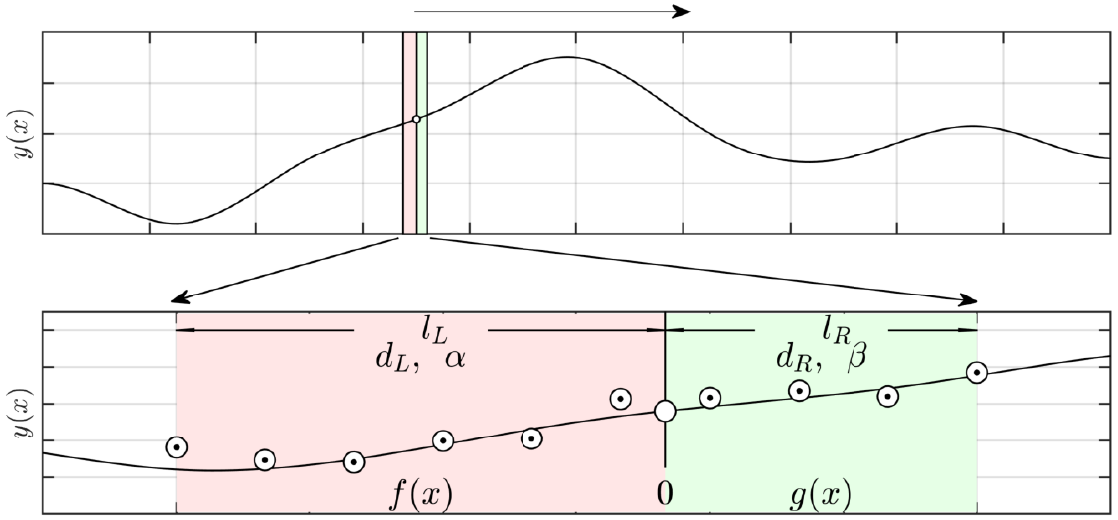}
	\caption{Schematic of a finite set of discrete observations (dotted circles) of a continuous function. The span of the observation is split into a left and right portion at the interstitial point (circle), with lengths $l_L$ and $l_R$ respectively. The left and right sides are considered to be the functions $f(x)$ and $g(x)$; modelled by the polynomials $\P{f}(x,\V{\alpha})$ and $\P{g}(x,\V{\beta})$ of degrees $d_L$ and $d_R$.}
	\label{fig:ContinuityDef}
\end{figure}

The basic concept (see Figure~\ref{fig:ContinuityDef}) to detect a $C^n$ discontinuity is: to approximate the data to the left of an interstitial point by the polynomial $\P{f}(x,\V{\alpha})$ of degree $d_L$ and to the right by $\P{g}(x,\V{\beta})$ of degree $d_R$, while constraining these approximations to be $C^{n-1}$ continuous at the interstitial point. This approximation ensures that,
\begin{equation}
\label{eqn.constraints}
\P{f}^{(k-1)}(\zeta_i) = \P{g}^{(k-1)}(\zeta_i),
\hspace{2ex}
\text{for every} \hspace{1ex} k\, \in \left[1 \ldots n\right].
\end{equation}
while yielding estimates for $\P{f}^{(n)}(\zeta_i)$ and $\P{g}^{(n)}(\zeta_i)$ together with estimates for their variances $\lambda_{f(\zeta_i)}$ and $\lambda_{g(\zeta_i)}$. This corresponds exactly to estimating the Taylor coefficients of the function twice for each interstitial point, i.e., once from the left and once from the right. It they differ significantly, then the function's $n^\text{th}$ derivative is discontinuous at this point.
The Taylor series of a function $f(x)$ around the point $a$ is defined as,
\begin{equation}
\label{eqn.Taylor}
f(x) = \sum_{k=0}^{\infty}\frac{f^{\left(k\right)}\left(a\right)}{k!}\left(x-a\right)^k
\end{equation}
for each $x$ for which the infinite series on the right hand side converges. Furthermore, any function which is $n+1$ times differentiable can be written as
\begin{equation}
f(x) = \tilde{\P{f}}(x) + R(x)
\end{equation}
where $\tilde{\P{f}}(x)$ is an $n^\text{th}$ degree polynomial approximation of the function $f(x)$,
\begin{equation}
\tilde{\P{f}}(x) = \sum_{k=0}^{n}\frac{f^{\left(k\right)}\left(a\right)}{k!}\left(x-a\right)^k
\end{equation}
and $R(x)$ is the remainder term. The Lagrange form of the remainder $R(x)$ is given by
\begin{equation}
R(x) = \frac{f^{\left(n+1\right)}\left(\xi\right)}{\left(n+1\right)!}\left(x-a\right)^{n+1}
\end{equation}
where $\xi$ is a real number between $a$ and $x$.

A Taylor expansion around the origin (i.e. $a = 0$ in Equation~\ref{eqn.Taylor}) is called a Maclaurin expansion; for more details, see~\cite{Burden1989}. In the rest of this work, the $n^\text{th}$ Maclaurin coefficient for the function $f(x)$ will be denoted by
\begin{equation}
t_{f}^{(n)} \defas \frac{f^{\left(n\right)}\left(0\right)}{n!}.
\end{equation}
The coefficients of a polynomial $\P{f}(x,\V{\alpha}) = \alpha_n x^n + ... +\alpha_1 x + \alpha_0$ are closely related to the coefficients of the Maclaurin expansion of this polynomial. Namely, it's easy to prove that
\begin{equation}
\alpha_k = t_\P{f}^{(k)}, \hspace{2ex}
\text{for every} \hspace{1ex} k\, \in \left[0 \ldots n\right].
\end{equation}
A prudent selection of a common local coordinate system, setting the interstitial point as the origin, ensures that the coefficients of the left and right approximating polynomials correspond to the derivative values at this interstitial point. Namely, one gets a very clear relationship between the coefficients of the left and right polynomial approximations, $\V{\alpha}$ and $\V{\beta}$, their Maclaurin coefficients, $t_{\P{f}}^{(n)}$ and $t_{\P{g}}^{(n)}$, and the values of the derivatives at the interstitial point
%
%
\begin{equation}
\label{eqn.TaylorVSPoly}
t_{\P{f}}^{(n)} = \alpha_n = \frac{\P{f}^{\left(n\right)}\left(0\right)}{n!}
\hspace{4ex}
\text{and}
\hspace{4ex}
t_{\P{g}}^{(n)} = \beta_n = \frac{\P{g}^{\left(n\right)}\left(0\right)}{n!}.
\end{equation}
From equation~\ref{eqn.TaylorVSPoly} it is clear that performing a left and right polynomial approximation at an interstitial point is sufficient to get the derivative values at that point, as well as their uncertainties.
\section{Constrained and Coupled Polynomial Approximation}
\label{CoupledPoly}
The goal here is to obtain $\Delta t_{\P{fg}}^{\left(n\right)} \defas t_{\P{f}}^{\left(n\right)} - t_{\P{g}}^{\left(n\right)}$ via polynomial approximation. To this end two polynomial approximations are required; whereby, the interstitial point is used as the origin in the common coordinate system, see Figure~\ref{fig:ContinuityDef}. The approximations are coupled~\cite{olearyDirect2006} at the interstitial point by constraining the coefficients such that $\alpha_i = \beta_i, \, \text{for every} \, i \in [0\ldots n-1]$. This ensures that the two polynomials are $C^{n-1}$ continuous at the interstitial points. This also reduces the degrees of freedom during the approximation and with this the variance of the solution is reduced. For more details on constrained polynomial approximation see~\cite{KlopfensteinConditional1964,olearyConstrained2019}.

To remain fully general, a local polynomial approximation of degree $d_L$ is performed to the left of the interstitial point with the support length $l_L$ creating $\P{f}(x,\V{\alpha})$; similarly to the right $d_R$, $l_R$, $\P{g}(x,\V{\beta})$. The $x$ coordinates to the left, denoted as $\V{x}_L$ are used to form the left Vandermonde matrix $\M{V}_L$, similarly  $\V{x}_R$ form $\M{V}_R$ to the right. This leads to the following formulation of the approximation process,
\begin{equation}    
\V{y}_L = \M{V}_L \, \V{\alpha}
\hspace{5mm}
\text{and}
\hspace{5mm}
\V{y}_R = \M{V}_R \, \V{\beta}.
\end{equation}
\begin{equation}
\begin{bmatrix}
\M{V}_L & \M{0} \\
\M{0} &  \M{V}_R
\end{bmatrix}
\,
\begin{bmatrix}
\V{\alpha} \\
\V{\beta}
\end{bmatrix}
=
\begin{bmatrix}
\V{y}_L \\
\V{y}_R
\end{bmatrix}
\end{equation}
A $C^{n-1}$ continuity implies $\alpha_i = \beta_i, \,\text{for every}\, i \in [0\ldots n-1]$ which can be written in matrix form as
\begin{equation}
\left[
\arraycolsep=4pt
\begin{array}{cc|cc}
\M{0}& \M{I}_{n-1}& \M{0}& - \M{I}_{n-1}
\end{array}
\right]
\,
\begin{bmatrix}
\V{\alpha} \\
\V{\beta}
\end{bmatrix}
=
\V{0}
\end{equation}
Defining
\begin{equation}\notag
\M{V} \defas
\begin{bmatrix}
\M{V}_L & \M{0} \\
\M{0} &  \M{V}_R
\end{bmatrix},
\,
\V{\gamma} \defas
\begin{bmatrix}
\V{\alpha} \\
\V{\beta}
\end{bmatrix},
\,
\V{y} \defas
\begin{bmatrix}
\V{y}_L \\
\V{y}_R
\end{bmatrix}
\,
\text{and}
\,
\M{C} \defas
\left[
\arraycolsep=2pt
\begin{array}{cc|cc}
\M{0}& \M{I}_{n-1}& \M{0}& - \M{I}_{n-1}
\end{array}
\right]
\end{equation}
We obtain the task of least squares minimization with homogeneous linear constraints,
\begin{empheq}[box=\widefbox]{align}
\min_{\V{\gamma}} \hspace{5mm} & \| \V{y} - \M{V} \, \V{\gamma} \|_2^2 \notag\\
\text{Given} \hspace{5mm} & \M{C} \, \V{\gamma} = \V{0}.\label{eqn.min1}
\end{empheq}
Clearly $\V{\gamma}$ must lie in the null-space of $\M{C}$; now, given $\M{N}$, an ortho-normal vector basis set for $\nullS{\M{C}}$, we obtain,
\begin{equation}
\V{\gamma} = \M{N} \, \V{\delta}.
\end{equation}
Back-substituting into Equation~\ref{eqn.min1} yields,
\begin{equation}
\min_{\V{\delta}} \| \V{y} - \M{V} \, \M{N} \, \V{\delta} \|_2^2
\end{equation}
The least squares solution to this problem is,
\begin{equation}
\V{\delta} = \left( \M{V} \, \M{N} \right)^+ \, \V{y},
\end{equation}
and consequently,
\begin{empheq}[box=\widefbox]{align}
\V{\gamma} = \begin{bmatrix}
\V{\alpha}\\
\V{\beta}
\end{bmatrix} = 
\M{N} \, \left( \M{V} \, \M{N} \right)^+ \, \V{y}
\end{empheq}
Formulating the approximation in the above manner ensures that the difference in the Taylor coefficients can be simply computed as 
\begin{equation}
\Delta t_{\P{fg}}^{\left(n\right)} = t_{\P{f}}^{\left(n\right)} - t_{\P{g}}^{\left(n\right)} = \alpha_n = \beta_n.
\end{equation}
Now defining $\V{d} = [1, \, \V{0}_{d_L -1}, \, -1, \, \V{0}_{d_R -1}]^\mathrm{T}$, $\Delta t_{\P{fg}}^{\left(n\right)}$ is obtained from $\V{\gamma}$ as 
\begin{equation}
\label{eqn.DeltaTaylor}
	\Delta t_{\P{fg}}^{\left(n\right)} = \VT{d}\V{\gamma} = \VT{d}\M{N} \, \left( \M{V} \, \M{N} \right)^+ \, \V{y}.
\end{equation}
%
%
\subsection{Covariance Propagation}
Defining, $\M{K} =  \M{N} \, \left( \M{V} \, \M{N} \right)^+$, yields,
$\V{\gamma} = \M{K} \, \V{y}$.
Then given the covariance of $\V{y}$, i.e., $\M{\Lambda}_{\V{y}}$, one gets that,
\begin{empheq}[box=\widefbox]{align}
\M{\Lambda}_{\V{\gamma}} = \M{K} \, \M{\Lambda}_{\V{y}} \, \MT{K}.
\end{empheq}
Additionally, from equation~\ref{eqn.DeltaTaylor} one could derive the covariance of the difference in the Taylor coefficients
\begin{equation}
\M{\Lambda}_{\V{\Delta}} = \V{d}\M{\Lambda}_{\V{\gamma}}\VT{d}
\end{equation}
Keep in mind that, if one uses approximating polynomials of degree $n$ to determine a discontinuity in the $n^{\text{th}}$ derivative, as done so far, $\M{\Lambda}_{\V{\Delta}}$ is just a scalar and corresponds to the variance of $\Delta t_{\P{fg}}^{\left(n\right)}$.
\section{Error Analysis}
\label{Residuals}
In this paper we consider three measures for error:
\begin{enumerate}
	\item the norm of the approximation residual;
	\item the combined approximation and extrapolation error;
	\item the extrapolation error.
\end{enumerate}
\subsection{Approximation Error}
The residual vector has the form
\begin{equation*}
\V{r} =\V{y}-\M{V}\V{\gamma} =\begin{bmatrix}
\V{y}_L-\M{V}_L\V{\alpha}\\
\V{y}_R-\M{V}_R\V{\beta}
\end{bmatrix}.  
\end{equation*}
The approximation error is calculated as
\begin{align*}
&E_a = \| \V{r}\|_2^2 = \| \V{y}_L-\M{V}_L\V{\alpha}\|_2^2 + \| \V{y}_R-\M{V}_R\V{\beta}\|_2^2\\
= &\left(\V{y}_L-\M{V}_L\V{\alpha}\right)^\mathrm{T}\left(\V{y}_L-\M{V}_L\V{\alpha}\right) + \left(\V{y}_R-\M{V}_R\V{\beta}\right)^\mathrm{T}\left(\V{y}_R-\M{V}_R\V{\beta}\right)\\
= &\VT{y}\V{y} - 2\VT{\alpha}\MT{V}_L\V{y}_L + \VT{\alpha}\MT{V}_L\M{V}_L\V{\alpha} - 2\VT{\beta}\MT{V}_R\V{y}_R + \VT{\beta}\MT{V}_R\M{V}_R\V{\beta}.
\end{align*}
\begin{figure}[t]
	\centering
	\includegraphics{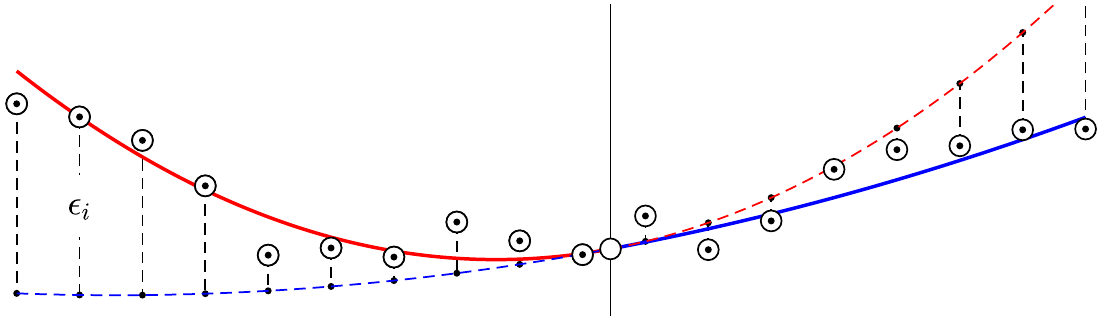}
	\caption{Schematic of the approximations around the interstitial point. Red: left polynomial approximation $\P{f}(x,\V{\alpha})$; dotted red: extrapolation of $\P{f}(x,\V{\alpha})$ to the RHS; blue: right polynomial approximation, $\P{g}(x,\V{\beta})$; dotted blue: extrapolation of $\P{g}(x,\V{\beta})$ to the LHS; $\epsilon_i$ is the vertical distance between the extrapolated value and the observation. The approximation is constrained with the conditions: $\P{f}(0,\V{\alpha}) = \P{g}(0,\V{\beta})$ and $\P{f}'(0,\V{\alpha}) = \P{g}'(0,\V{\beta})$.}
	\label{fig:taylorPolyApprox}
\end{figure}
\subsection{Combined Error}
The basic concept, which can be seen in Figure~\ref{fig:taylorPolyApprox}, is as follows: the left polynomial $\P{f}\left(x,\V{\alpha}\right)$, which approximates over the values $\V{x}_L$, is extended to the right and evaluated at the points $\V{x}_R$. Analogously, the right polynomial $\P{g}\left(x,\V{\beta}\right)$ is evaluated at the points $\V{x}_L$. If there is no $C^n$ discontinuity in the system, the polynomials $\P{f}$ and $\P{g}$ must be equal and consequently the extrapolated values won't differ significantly from the approximated values.
\subsubsection{Analytical Combined Error}
The extrapolation error in a continuous case, i.e. between the two polynomial models, can be computed with the following $2$-norm,
\begin{equation}
\epsilon_x =
\int_{x_{min}}^{x_{max}}
\left\{ \P{f}(x,\V{\alpha}) - \P{g}(x,\V{\beta})
\right\}^2 \, \md x.
\end{equation}
Given, the constraints which ensure that $\alpha_i = \beta_i \, i \in [0,\ldots,n-1]$, we obtain,
\begin{equation}
\epsilon_x =
\int_{x_{min}}^{x_{max}}
\left\{ (\alpha_{n} - \beta_{n} ) \, x^{n}
\right\}^2 \, \md x.
\end{equation}
Expanding and performing the integral yields,
\begin{equation}
\epsilon_x = (\alpha_{n} - \beta_{n})^2 \, \left\{
\frac{x_{max}^{2n + 1} - x_{min}^{2n + 1}}
{2n + 1} \right\}
\end{equation}
Given fixed values for $x_{min}$ and $x_{max}$ across a single computation implies that the factor,
\begin{equation}
k = 
\frac{x_{max}^{2n + 1} - x_{min}^{2n + 1}}
{2n + 1} 
\end{equation}
is a constant. Consequently, the extrapolation error is directly proportional to the square of the difference in the Taylor coefficients,
\begin{equation}
\epsilon_x \propto \, \left(\alpha_n - \beta_n\right)^2 \propto \, \left\{\Delta t_{\P{fg}}^{\left(n\right)}\right\}^2.
\end{equation}
\subsubsection{Numerical Combined Error}
In the discrete case, one can write the errors of $\P{f}(x,\V{\alpha})$ and $\P{g}(x,\V{\beta})$ as 
\begin{equation}
	\V{e}_\P{f} = \V{y}-\P{f}(\V{x},\V{\alpha}) \hspace{2ex} \text{and} \hspace{2ex}
	\V{e}_\P{g} = \V{y}-\P{g}(\V{x},\V{\beta})
\end{equation}
respectively. Consequently, one could define an error function as
\begin{align}
	&E_{\P{f}\P{g}} = \|\V{e}_\P{f}-\V{e}_\P{g}\|_2^2 =
	\|(a_{n} - b_{n} ) \, \V{z}\|_2^2 = 
	(a_{n} - b_{n} )^2 \VT{z} \V{z}^{n} =
	(a_{n} - b_{n} )^2\sum {x_i^{n}}
\end{align}
where $\V{z} \defas \V{x} .\, \hat{} \,\, n$. From these calculations it is clear that in the discrete case the error is also directly proportional to the square of the difference in the Taylor coefficients and that $ E_{\P{f}\P{g}} \propto \epsilon_x$. This proves that the numerical computation is consistent with the analytical continuous error.
\subsection{Extrapolation Error}
One could also define a different kind of error, based just on the extrapolative properties of the polynomials. Namely, using the notation from the beginning of Section~\ref{CoupledPoly}, one defines
\begin{equation*}
\V{r}_{e\P{f}} = \V{y}_L-\P{g}(\V{x}_L,\V{\beta}) = \V{y}_L-\M{V}_L\V{\beta} 
\hspace{2ex} \text{and} \hspace{2ex}
\V{r}_{e\P{g}} = \V{y}_R-\P{f}(\V{x}_R,\V{\alpha}) = \V{y}_R-\M{V}_R\V{\alpha}
\end{equation*}
and then calculates the error as
\begin{align*}
&E_{e} = \VT{r}_{e\P{f}}\V{r}_{e\P{f}} + \VT{r}_{e\P{g}}\V{r}_{e\P{g}}\\
=&\left(\V{y}_L-\M{V}_L\V{\beta}\right)^\mathrm{T} \left(\V{y}_L-\M{V}_L\V{\beta}\right) + \left(\V{y}_R-\M{V}_R\V{\alpha}\right)^\mathrm{T} \left(\V{y}_R-\M{V}_R\V{\alpha}\right)\\
=&\VT{y}\V{y} -2\VT{\beta}\MT{V}_L \V{y}_L + \VT{\beta}\MT{V}_L\M{V}_L\V{\beta} - 2\VT{\alpha}\MT{V}_R\V{y}_R + \VT{\alpha}\MT{V}_R\M{V}_R\V{\alpha}.
\end{align*}
In the example in section~\ref{sec.examples}, it will be seen that there is no significant numerical difference between these two errors.
\section{Numerical Testing}
\label{sec.examples}
The numerical testing is performed with: synthetic data from a piecewise polynomial, where the locations of the $C^n$ discontinuities are known; and with real sensor data emanating from the monitoring of heavy machinery.
\subsection{Synthetic Data}
\begin{figure}[h]
	\centering
	\includegraphics{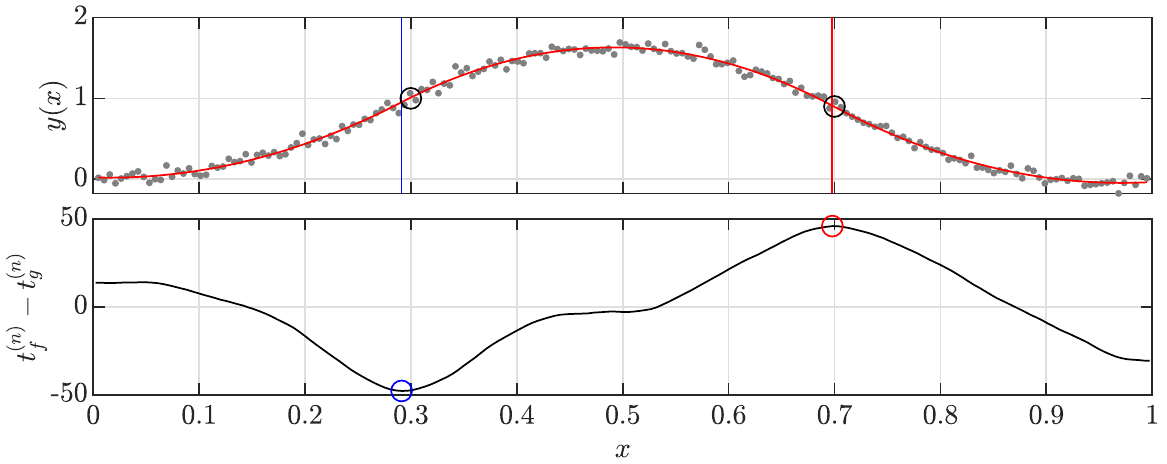}
	\caption{A piecewise polynomial of degree $d=2$, created from the 
		knots sequence $\V{x}_k = [0, 0.3, 0.7, 1]$ with the corresponding 
		values $\V{y}_k = [0, 0.3, 0.7, 1]$. The end points are clamped with 
		$y'(x)_{0,1} = 0$. Gaussian noise is added with $\sigma = 0.05$. Top: 
		the circles mark the known points of $C^2$ discontinuity; the blue and 
		red lines indicate the detected discontinuities; additionally the data 
		has been approximated by the b-spline (red) using the detected 
		discontinuities as knots. Bottom: shows $\Delta t^{(n)}_{\P{fg}} = t^{(n)}_\P{f} - 
		t^{(n)}_\P{g}$, together with the two identified peaks.}
	\label{PieceWisePoly}
\end{figure}

In the literature on splines, functions of the type $y\left(x\right) = e^{-x^2}$ are commonly used. However, this function is analytic and $C^{\infty}$ continuous; consequently it was not considered a suitable function for testing. In Figure~\ref{PieceWisePoly} a piecewise polynomial with a similar shape is shown; however, this curve has $C^2$ discontinuities at known locations.
The algorithm was applied to the synthetic data from the piecewise polynomial, with added noise with $\sigma = 0.05$ and the results for a single case can be seen in Figure~\ref{PieceWisePoly}. Additionally, a Monte Carlo simulation with $m=10000$ iterations was performed and the results of the algorithm were compared to the true locations of the two known knots. The mean errors in the location of the knots are: $\mu_1 = (5.59 \pm 
2.05) \times 10^{-4}$ with $95 \%$ confidence, and $\mu_2 = (-4.62 \pm 1.94)\times 10^{-4}$. Errors in the scale of $10^{-4}$, in a support with a range $[0,\,1]$, and $5 \%$ noise amplitude in the curve can be considered a highly satisfactory result.
\subsection{Sensor Data}
\begin{figure}[h]
	\centering
	\includegraphics{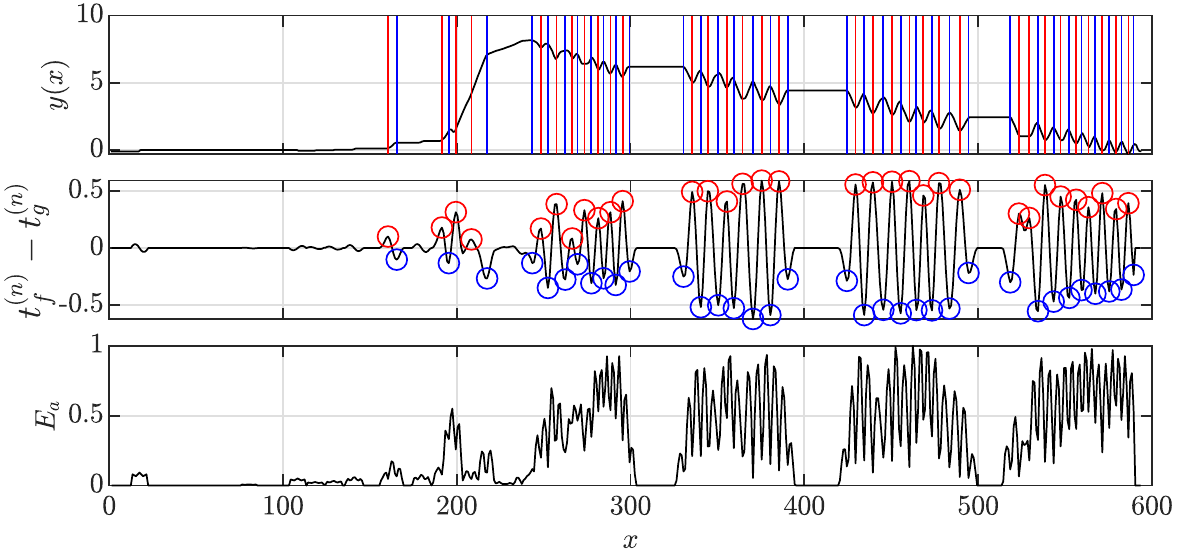}
	\caption{The top-most graph shows a function $y(x)$, together with the detected $C^1$ discontinuity points. The middle graph shows the difference in the Taylor polynomials $\Delta t_{\P{fg}}^{\left(n\right)}$ calculated at every interstitial point. The red and blue circles mark the relevant local maxima and minima of the difference respectively. According to this, the red and blue lines are drawn in the top-most graph. The bottom graph shows the approximation error evaluated at every interstitial point.}
	\label{fig:taylorCnDisKeller}
\end{figure}

The algorithm was also applied to a set of real-world sensor data\footnote{For confidentiality reasons the data has been anonymized.} emanating from the monitoring of heavy machinery. The original data set can be seen in Figure~\ref{fig:taylorCnDisKeller} (top). It has many local peaks and periods of little or no change, so the algorithm was used to detect discontinuities in the first derivative, in order to determine the peaks and phases.
\begin{figure}[h]
	\centering
	\includegraphics{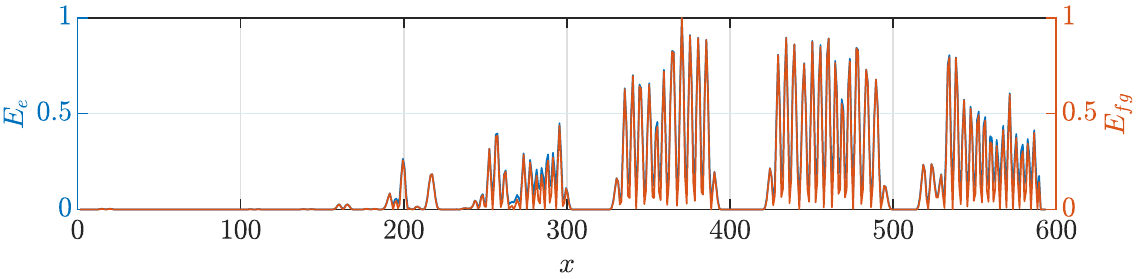}
	\caption{The two error functions, $E_e$ and $E_{\P{f}\P{g}}$ as defined in Section \ref{Residuals}, for the example from Fig.~\ref{fig:taylorCnDisKeller}. One can see that the location of the peaks doesn't change, and the two errors don't differ significantly.}
	\label{fig:taylorCnDisKellerErrors}
\end{figure}
The peaks in the Taylor differences were used in combination with the peaks of the extrapolation error to determine the points of discontinuity. A peak in the Taylor differences means that the Taylor coefficients are significantly different at that interstitial point, compared to other interstitial points in the neighbourhood. However, if there is no peak in the extrapolation errors at the same location, then the peak found by the Taylor differences is deemed insignificant, since one polynomial could model both the left and right values and as such the peak isn't a discontinuity.
Additionally, it can be seen in Figure~\ref{fig:taylorCnDisKellerErrors} that both the extrapolation error and the combined error, as defined in Section~\ref{Residuals}, have peaks at the same locations, and as such the results they provide do not differ significantly.
\section{Conclusion and Future Work}
It may be concluded, from the results achieved, that the coupled constrained polynomial approximation yield a good method for the detection of $C^n$ discontinuities in discrete observational data of continuous systems.
Local peaks in the square of the difference of the Taylor polynomials provide a relative measure as a means of determining the locations of discontinuities.

Current investigations indicate that the method can be implemented directly as a convolutional operator, which will yield a computationally efficient solution. The use of discrete orthogonal polynomials~\cite{Persson2003Smoothing,oleary2010Discrete} is being tested as a means of improving the sensitivity of the results to numerical perturbations.
\section*{Acknowledgements}
This work was partially funded by:
\begin{enumerate}
	\item The COMET program within the K2 Center “Integrated Computational Material, Process and Product Engineering (IC-MPPE)” (Project No 859480). This program is supported by the Austrian Federal Ministries for Transport, Innovation and Technology (BMVIT) and for Digital and Economic Affairs (BMDW), represented by the Austrian research funding association (FFG), and the federal states of Styria, Upper Austria and Tyrol.
	\item The European Institute of Innovation and Technology (EIT), a body of the European Union which receives support from the European Union's Horizon 2020 research and innovation programme. This was carried out under Framework Partnership Agreement No. 17031 (MaMMa - Maintained Mine \& Machine).
\end{enumerate}
The authors gratefully acknowledge this financial support.

\bibliographystyle{splncs04}
\bibliography{bibliography}
\end{document}